# PHASE–TRANSITION THEORY OF INSTABILITIES.
# IV.  CRITICAL POINTS ON THE MACLAURIN SEQUENCE AND NONLINEAR FISSION PROCESSES

DIMITRIS M. CHRISTODOULOU[1], DEMOSTHENES KAZANAS[2], ISAAC SHLOSMAN[3,4], AND JOEL E. TOHLINE[5]

## ABSTRACT

We use a free–energy minimization approach to describe in simple and clear physical terms the secular and dynamical instabilities as well as the bifurcations along equilibrium sequences of rotating, self–gravitating fluid systems. Our approach is fully nonlinear and stems from the Landau–Ginzburg theory of phase transitions. In the final paper of this series, we examine higher than second–harmonic disturbances applied to Maclaurin spheroids, the corresponding bifurcating sequences, and their relation to nonlinear fission processes.

The triangle and ammonite sequences bifurcate from the two third–harmonic neutral points on the Maclaurin sequence while the square and one–ring sequences bifurcate from two of the three known fourth–harmonic neutral points. The one–ring sequence has been analyzed in Paper II. In the other three cases, secular instability does not set in at the corresponding bifurcation points because the sequences stand and terminate at higher energies relative to the Maclaurin sequence. Consequently, an anticipated (numerically unresolved) third–order phase transition at the ammonite bifurcation and numerically resolved second–order phase transitions at the triangle and square bifurcations are strictly forbidden. Furthermore, the ammonite sequence exists at higher rotation frequencies as well and is similar in every respect to the pear–shaped sequence that has been analyzed in Paper III.

There is no known bifurcating sequence at the point of third–harmonic dynamical instability. This point represents a discontinuous $\lambda$–transition of type 3 that brings a

---

[1]Virginia Institute for Theoretical Astronomy, Department of Astronomy, University of Virginia, P.O. Box 3818, Charlottesville, VA 22903

[2]NASA Goddard Space Flight Center, Code 665, Greenbelt, MD 20771

[3]Gauss Foundation Fellow

[4]Department of Physics & Astronomy, University of Kentucky, Lexington, KY 40506

[5]Department of Physics & Astronomy, Louisiana State University, Baton Rouge, LA 70803

Maclaurin spheroid on a dynamical time scale directly to the binary sequence while the original symmetry and topology are broken in series. The remaining fourth–harmonic neutral point also appears to be related to a type 3 $\lambda$–transition which however takes place from the lower turning point of the one–ring sequence toward the starting point and then on toward the stable branch of the three–fluid–body (triple) sequence. A third type 3 $\lambda$–transition, taking place from the one–ring sequence toward the starting point and then on toward the stable branch of the four–fluid–body (quadruple) sequence, is also discussed.

The two–ring sequence bifurcates from the axisymmetric sixth–harmonic neutral point on the Maclaurin sequence also toward higher energies initially but eventually turns around and proceeds to lower energies relative to the Maclaurin sequence. The point where the two sequences have equal energies represents a fourth type of $\lambda$–transition which is not preceded by a first–order phase transition. This type 4 $\lambda$–transition results in double fission on a secular time scale: a Maclaurin spheroid breaks into two coaxial axisymmetric tori that rotate uniformly and with the same frequency.

Finally, our nonlinear approach easily identifies *resonances* between the Maclaurin sequence and various multi–fluid–body sequences that cannot be detected by linear stability analyses. Resonances appear as first–order phase transitions at points where the energies of the two sequences are nearly equal but the lower energy state belongs to one of the multi–fluid–body sequences. Three nonlinear resonances leading to the turning points of the binary, triple, and quadruple sequences are described.

*Subject headings:*   galaxies: evolution – galaxies: structure – hydrodynamics – instabilities – stars: formation

## 1    INTRODUCTION

In this series of papers, we explore the physics of bifurcation and instability points along equilibrium sequences of uniformly rotating, self–gravitating, incompressible fluids using a method that stems from the Landau–Ginzburg theory of phase transitions (see Landau & Lifshitz 1986; see also related works by Bertin & Radicati 1976, Lebovitz 1977, Poston & Stewart 1978, Constantinescu, Michel, & Radicati 1979, Gilmore 1981, Hachisu & Eriguchi 1983, 1984a, Eriguchi & Hachisu 1983b, 1984, Tohline 1985, and Christodoulou & Tohline 1990). In the first three papers (hereafter referred to as Papers I–III), we have described the differences between the second–order phase transitions that correspond to the second–harmonic instabilities on the Maclaurin sequence (Paper I), we have investigated the first–order phase transitions and the discontinuous $\lambda$–transitions related to the one–ring, dumbbell–binary, and Maclaurin toroid sequences (Paper II), and we have analyzed the third–harmonic bifurcation of the pear–shaped sequence from the Jacobi sequence and its relation to the classical fission hypothesis (Paper III).

Phase–transition theory allows us to understand all points of bifurcation and instability along equilibrium sequences as points where phase transitions of various orders may appear and may either be *allowed* or may be *forbidden* depending on the (non)–conservation of four "integrals of motion." These integrals of motion are the mass, the energy, the angular momentum, and the circulation.

The most important contribution of this series of papers is the realization that the above integrals of motion determine alone whether/where points of secular and dynamical instability exist. In our terminology, integrals of motion determine whether a phase transition appears at a point and whether a known phase transition is allowed or forbidden. This means that an accurate description of an evolutionary path requires knowledge of whether any integral of motion varies during evolution and understanding of how such an integral may vary in time. Consider two examples that are particularly relevant to the results described below. (1) In general, circulation is not conserved between sequences of uniformly rotating equilibria and poses no constraint to the occurence of the corresponding phase transitions. Even so, not every neutral point is a point of secular instability because a bifurcating "daughter" sequence may unfold, at least initially, toward higher energies relative to its "mother" sequence (cf. Papers II and III). (2) The instabilities/transitions detected by linear stability analyses are not the only allowed phase transitions between sequences. In particular, linear analyses do not capture nonlinear resonances between sequences (see below), first–order phase transitions that represent nonlocal changes, or type 2 $\lambda$–transitions that represent discontinuous catastrophes (see Paper II).

In this paper, we complete our analysis of the known critical points and the corresponding bifurcating sequences of the Maclaurin sequence. Specifically, we describe the phase transitions that are related to: (a) the two third–harmonic neutral points and the corresponding bifurcating (triangle and ammonite) sequences; (b) the third–harmonic dynamical instability point; (c) the three fourth–harmonic neutral points and the two known bifurcating (square and one–ring) sequences; (d) the axisymmetric sixth–harmonic bifurcation of the two–ring sequence; and (e) nonlinear resonances between the Maclaurin and various multi–fluid–body (binary, triple, quadruple) sequences.

In what follows, we consider for simplicity only the evolution of incompressible, uniformly rotating fluid–masses that collapse under the action of self–gravity. We also adopt again the assumptions made in Papers II and III: (1) Mass is always strictly conserved not only between the initial state and the final equilibrium state but also in the intermediate states that an evolving fluid–mass passes through. (2) Energy always decreases as a fluid–mass evolves toward a new equilibrium state. (3) Angular momentum, expressed in physical units, is always strictly conserved but, in normalized units, it increases or stays constant in time. This general increase in normalized angular momentum $j$ reflects a general increase of the density $\rho$ during the evolution of a contracting fluid–mass [$j \propto \rho^{1/6}$ in equation (1.2) below]. (4) Circulation is strictly conserved only in the complete absence of "viscosity" (i.e. in a "perfect" fluid) but its absolute value decreases during the evolution of a "viscous" fluid–mass.

Our interpretations of instability points rely on numerical results obtained by Eriguchi, Hachisu, and Sugimoto between 1981 and 1986. We refer below to particular results from these numerical computations as we need them. We also adopt for convenience the dimensionless quantities $\omega$ (rotation frequency), $j$ (angular momentum), and $E$ (total energy) that were used by these authors as well as in Papers II and III. These quantities are defined by

$$\omega^2 \equiv 10^2 \Big(\frac{\Omega^2}{4\pi G\rho}\Big), \tag{1.1}$$

$$j^2 \equiv 10^2 \Big(\frac{L^2}{4\pi G\rho^{-1/3}M^{10/3}}\Big), \tag{1.2}$$

and

$$E \equiv 10^4 \Big[\frac{T+W}{(4\pi G)^2 M^5 L^{-2}}\Big], \tag{1.3}$$

where $\Omega$ is the rotation frequency, $G$ is the gravitational constant, $\rho$ is the mass density, $L$ is the angular momentum, and $M$ is the mass, all in physical units. The term $T+W$ denotes the total energy in physical units as the sum of the total kinetic energy due to rotation $T$ and the total gravitational potential energy $W$. As we have seen in Paper I, the total energy $T+W$ and its dimensionless counterpart $E$ also represent the free-energy function when they are not constrained by equilibrium conditions (for more details see Tohline & Christodoulou 1988). Finally, when $\omega$, $j$, and $E$ refer to equilibrium configurations specifically we write them as $\omega_o$, $j_o$, and $E_o$, respectively.

The remainder of the paper is organized as follows. In §2, we describe the triangle and ammonite sequences that bifurcate from the two third–harmonic neutral points on the Maclaurin sequence and the square sequence that bifurcates from one of the fourth–harmonic neutral points. These three neutral points are not points of secular instability. The second–order phase transitions toward the triangle and square sequences are forbidden because these sequences exist only at higher energies relative to the Maclaurin sequence. The ammonite sequence is also unstable and is further similar to the pear–shaped sequence that was analyzed in Paper III. All three sequences terminate prematurely because of equatorial mass shedding. Consequently, there exist no related first–order phase transitions or $\lambda$–transitions similar to those found in Paper II to be associated with the one–ring sequence. In §3, we describe the two-ring sequence that bifurcates from the axisymmetric sixth–harmonic neutral point on the Maclaurin sequence. This sequence initially unfolds toward higher energies but eventually crosses below the Maclaurin sequence. A new type of symmetry–preserving, topology–breaking $\lambda$–transition appears at the point where the two sequences have equal energies. Because of the particular structure of the two–ring sequence, this type 4 $\lambda$–transition is not preceded by a first–order phase transition.

In §4, we describe the $\lambda$–transitions of type 3 that appear to be associated with the third–harmonic dynamical instability point and with two of the fourth–harmonic neutral points on the Maclaurin sequence. The latter two transitions originate from points on the lower energy toroidal branch of the one–ring sequence. Type 3 $\lambda$–transitions lead to nonlinear fission of a Maclaurin spheroid or of an axisymmetric torus to multiple fluid bodies while first the symmetry and then the topology of the original figure are broken. The initial breaking of the symmetry (which is characteristic of only type 3 $\lambda$–transitions) must be the reason that this type of dynamical instability is detected by linear stability analyses (see Chandrasekhar 1969, hereafter referred to as EFE). In §5, we describe the nonlinear resonances that exist between the Maclaurin sequence and various multi–fluid–body sequences. Resonances are related to the turning points of the binary, triple, and quadruple sequences and appear as first–order phase transitions that also result in nonlinear fission for sufficiently strong disturbances. In §6, we summarize our results and we relate them to previous linear stability analyses and numerical hydrodynamical simulations.

FIGURE 1. The rotation frequency squared $\omega_o^2$ (left panels) and the free energy $E_o$ (right panels) along the triangle, square, ammonite, and Maclaurin sequences are plotted as functions of the dimensionless angular momentun squared $j_o^2$. The Maclaurin sequence and the exact locations of the bifurcation points B (solid circles) have been calculated analytically (see also Table 1). All points (open circles) of the bifurcating sequences were obtained from Tables II–IV of Eriguchi & Hachisu (1982) who gave the points denoted by the letter T as the termination points of the sequences due to equatorial mass shedding.

## 2 The Triangle, Square, and Ammonite Sequences

The triangle, square, and ammonite sequences bifurcate from the Maclaurin sequence at the neutral points that correspond to the harmonic indices $(l,m)$=(3,3), (4,4), and (3,1), respectively (see EFE and Hachisu & Eriguchi 1984c). The physical properties of these sequences were computed by Eriguchi & Hachisu (1982). Their results are shown in Figure 1 where the letters B denote the exact locations of the bifurcation points (EFE) and the letters T denote the termination points of the sequences due to mass shedding. In the $(j_o^2, E_o)$ plane, the triangle sequence terminates above but close to the Maclaurin sequence while the square sequence turns around toward lower $j_o$–values and slightly higher $E_o$–values before terminating.

We see then from Figure 1 that there is no instability of any kind related to the triangle and square sequences. Consequently, the phase transitions at the bifurcation points of these sequences are strictly forbidden. Notice, however, that the slopes $dE_o/dj_o$ appear to be continuous while the slopes $d\omega_o/dj_o$ have a finite discontinuity at the bifurcation points. (This conclusion is certainly valid despite the mismatch seen at the bifurcation points in Figure 1 between the analytical and the numerical results.) This is consistent with our description of second–order phase transitions given in Paper I.

The ammonite sequence is different than both the above sequences and similar in every detail to the pear–shaped sequence that bifurcates from the Jacobi sequence (Paper III). In this case as well, the sequence exists only at higher rotation frequencies and at higher energies relative to the Maclaurin sequence and no instability of any kind appears. We believe then that the ammonite sequence bifurcates in the $(j_o^2, \omega_o^2)$ plane smoothly toward lower $j_o$–values and higher $\omega_o$–values initially and that the bifurcation point represents a forbidden third–order phase transition (compare the relevant panels in Figure 1 to the two figures in Paper III). This conclusion about the overall structure of the ammonite sequence is identical to the results of Paper III concerning the (3,1) neutral point on the Jacobi sequence and the behavior of the corresponding bifurcating pear–shaped sequence.

For convenience, we list in Table 1 the physical properties of the above discussed neutral points as well as of the critical points discussed below. The meridional eccentricity $e$ of the corresponding Maclaurin spheroid at each point is also listed along with the values of $j_o^2$, $\omega_o^2$, $E_o$, and the ratio $T/|W|$. The second–harmonic instability points discussed in Paper I are also listed in Table 1 for completeness. The first entries of the (4,0) neutral point were calculated using the value $e$=0.98531 (Chandrasekhar 1968). The second entries of the (4,0) neutral point are taken from Paper II (they were obtained from the results of Chandrasekhar 1967) and are listed here for comparison purposes. The numerical differences between these two determinations are practically negligible.

TABLE 1
CRITICAL POINTS ON THE MACLAURIN SEQUENCE

| Point $(l,m)$ | Bifurcating Sequence | $e$ | $j_o^2$ | $\omega_o^2$ | $E_o$ | $T/|W|$ |
|---|---|---|---|---|---|---|
| \multicolumn{7}{c}{SECOND–HARMONIC POINTS} |
| Neutral (2,2) | Jacobi | 0.81267 | 0.45547 | 9.35574 | −2.94820 | 0.13753 |
| Dynamical Instability[a] | $x = +1$ Riemann | 0.95289 | 1.27957 | 11.00550 | −6.36714 | 0.27382 |
| \multicolumn{7}{c}{THIRD–HARMONIC POINTS} |
| Neutral (3,3) | Triangle | 0.89926 | 0.78526 | 11.00366 | −4.55015 | 0.20233 |
| Dynamical Instability[b] | ... | 0.96696 | 1.53796 | 10.49141 | −7.10294 | 0.30308 |
| Neutral (3,1) | Ammonite | 0.96937 | 1.59504 | 10.35347 | −7.24912 | 0.30895 |
| \multicolumn{7}{c}{FOURTH–HARMONIC POINTS} |
| Neutral (4,4) | Square | 0.93275 | 1.03712 | 11.23066 | −5.55125 | 0.24174 |
| Neutral (4,2) | ... | 0.98097 | 1.96741 | 9.33478 | −8.07828 | 0.34290 |
| Neutral (4,0) | One–ring | 0.98531 | 2.17884 | 8.71210 | −8.46563 | 0.35925 |
| Neutral (4,0) | One–ring (Paper II) | 0.98523 | 2.17412 | 8.72609 | −8.45757 | 0.35891 |
| \multicolumn{7}{c}{SIXTH–HARMONIC POINTS} |
| Neutral (6,0) | Two–ring | 0.99375 | 2.91500 | 6.61108 | −9.46087 | 0.40344 |
| Secular Instability[c] | ... | 0.99802 | 3.98856 | 4.21544 | −10.27277 | 0.44322 |

(a) Also critical point of a second–order phase transition (Paper I).
(b) Also a type 3 $\lambda$–point (§4).
(c) Also a type 4 $\lambda$–point (§3).

## 3    THE TWO–RING SEQUENCE

The two–ring sequence is shown in Figures 2 and 3; it bifurcates from the Maclaurin sequence at the (6,0) neutral point that corresponds to a meridional eccentricity of $e$=0.99375 (Hachisu & Eriguchi 1984c). The physical properties of this sequence were computed by Eriguchi & Hachisu (1982, 1983a). Like the one–ring sequence of Paper II, the two–ring sequence bifurcates smoothly in both Figures 2 and 3 and unfolds toward higher energies initially. This implies that the third–order phase transition at the bifurcation point is forbidden as in the case of the one–ring sequence.

FIGURE 2. The rotation frequency squared $\omega_o^2$ along the two–ring and the Maclaurin sequence is plotted as a function of the dimensionless angular momentun squared $j_o^2$. The Maclaurin sequence has been calculated analytically. The bifurcation point B (filled circle) has also been calculated analytically using an angular momentum value of $j_o^2 = 2.915$ (Hachisu & Eriguchi 1984c and Table 1). All points (open circles) of the two–ring sequence other than the bifurcation point were obtained from the corresponding tables of Eriguchi & Hachisu (1982, 1983a). The dotted line XC at $j_o^2 = 3.98856$ denotes a type 4 $\lambda$–transition (secular instability) that is not preceded by a first–order phase transition.

In contrast to the one–ring sequence, four turning points can be seen in the two–ring sequence of Figure 2. These turning points are associated with changes in the structure of the equilibrium objects (Eriguchi & Hachisu 1983a) and cause a difference in the behavior of the sequence in the $(j_o^2, E_o)$ plane of Figure 3: The two additional turning points make the sequence turn forward toward higher $j_o$–values. The two–ring sequence eventually crosses below the Maclaurin sequence but a first–order phase transition similar to that of the one–ring sequence in Paper II is not realized because there is no intermediate branch to act as a free–energy barrier at the point where the two sequences have equal free energies. Thus, this point is a degenerate critical point of the free–energy function.

FIGURE 3. *The free energy $E_o$ is plotted versus $j_o^2$ for the sequences of Figure 2. The intersection X=C between the Maclaurin and the two–ring sequence is a $\lambda$–point of type 4 (see also Table 1).*

The degenerate critical point (X=C) between the two sequences in Figure 3 occurs at $j_o^2 = 3.98856$, $E_o = -10.27277$ and corresponds to point X on the Maclaurin sequence of Figure 2 with $e = 0.99802$, $\omega_o^2 = 4.21544$, and $T/|W| = 0.44322$. This point is a $\lambda$–point of a new type which we call type 4. The type 4 $\lambda$–transition from point X to point C in Figure 2 results in axisymmetric nonlinear fission of the original Maclaurin spheroid to two toroidal figures that rotate uniformly and with the same frequency. Thus, this transition may only take place on a secular time scale in the presence of viscosity. The axisymmetry of the original figure is preserved but the topology (i.e. the connectedness of volume) is broken. The two tori along the lower branch of the two–ring sequence have inner–to–outer mass ratio of about 0.9. Other systems of two tori with different mass ratios do not bifurcate from the Maclaurin sequence; they seem to connect to various core–ring sequences (Eriguchi & Hachisu 1983a).

Neither the bifurcation point nor the $\lambda$–point of the two–ring sequence (Table 1) is related to the axisymmetric dynamical instability found by Bardeen (1971) at $e = 0.998917$, $j_o^2 = 4.598734$, $\omega_o^2 = 3.24702$, $E_o = -10.53929$, and $T/|W| = 0.457424$. This reinforces our conclusion in Paper II that Bardeen's dynamical instability represents a type 1 $\lambda$–transition from the bifurcation point of the Maclaurin toroid sequence to the stable lower energy branch of the same sequence.

In its simplest level, the type 4 $\lambda$–transition of the two–ring sequence is part of the "cusp catastrophe." Prior to the intersection X=C in Figure 3, the Maclaurin line represents stable equilibria while the two–ring branches represent maxima or inflection points of higher free energies. Such maxima and inflection points can only be explained at a higher level as parts of a higher elementary catastrophe. However, near point X=C, the free–energy diagram is simplified considerably. The two sequences exhibit degeneracy and their roles are reversed past this point. It is this smooth exchange of roles across X=C that can be described by an evolutionary path in the control plane of the cusp catastrophe. This path lies outside the fold lines that characterize doubly degenerate critical points (see Figure 5.4 and §5.4 in Gilmore 1981). Staying outside the fold lines eliminates the first–order phase transition encountered prior to type 1 and type 2 $\lambda$–transitions (Paper II).

In summary, the above type 4 $\lambda$–transition differs from the type 1 $\lambda$–transition of Paper II in four respects: (a) it does not exhibit a first–order phase transition prior to the $\lambda$–point; (b) it represents a secular instability; (c) the "specific heat" diverges at the $\lambda$–point; and (d) it results in "double fission" to two toroidal figures. On the other hand, both types of transitions share several common properties since axisymmetry is preserved, topology breaks, and both evolutionary paths are parts of the cusp catastrophe (see also Table 3 below).

## 4    Type 3 $\lambda$–transitions

Third–harmonic dynamical instability sets in at $e$=0.96696 on the Maclaurin sequence (EFE) but there is no known sequence that bifurcates at that point. As we have briefly discussed in Paper II, the point of third–harmonic dynamical instability represents a type 3 $\lambda$–transition between the Maclaurin sequence and the lower energy stable branch of the dumbbell–binary sequence. This $\lambda$–transition is shown in Figures 4 and 5 along with two more $\lambda$–transitions of the same type that originate on the lower branch of the one–ring sequence.

We have determined the properties of the type 3 $\lambda$–points based on the $j_o$–values of the turning or starting points of the corresponding sequences. Analytical values of the relevant critical points on the Maclaurin sequence were obtained from Chandrasekhar (1968) and from EFE. Our results are summarized also in Table 2:

(a) The $\lambda$–transition M$\rightarrow$ D$\rightarrow$ B in Figures 4 and 5 was determined from the value $j_o^2 = 1.540$ at the higher turning point D of the dumbbell–binary sequence. The $\lambda$–point M with the same $j_o$–value on the Maclaurin sequence is identical to the point of third–harmonic dynamical instability (M$'$ in Table 2). The transition leads to a detached binary (point B in Table 2 and in Figures 4 and 5).

FIGURE 4. The rotation frequency squared $\omega_o^2$ along several incompressible sequences is plotted as a function of the dimensionless angular momentun squared $j_o^2$. The one–ring sequence was obtained from Table I of Eriguchi & Sugimoto (1981). The dumbbell–binary sequence was obtained from Table II of Eriguchi, Hachisu, & Sugimoto (1982) and from Table 1 of Hachisu & Eriguchi (1984a). All points (open circles) of the triple and quadruple sequences were obtained from Table I of Eriguchi & Hachisu (1983b). The Maclaurin and Jacobi sequences and the bifurcation points (squares) of the dumbbell–binary and one–ring sequences (see Paper II) were calculated analytically. Point M lies directly above the higher turning point D of the dumbbell sequence where $j_o^2$=1.540. The point of third–harmonic dynamical instability on the Maclaurin sequence with $j_o^2$=1.53796 (corresponding to a meridional eccentricity of $e$=0.96696; EFE) coincides with M. Points $M_3$ and $M_4$, corresponding to $e$=0.98126 and 0.98495, lie directly above points $R_3$ and $S_4$ (and $R_4$), respectively. They were determined using the values $j_o^2$=1.980 (lower turning point $R_3$ of the one–ring sequence) and $j_o^2$=2.159 (starting point $S_4$ of the quadruple sequence), respectively. Two of the three known fourth–harmonic neutral points (corresponding to $e$=0.98097 and 0.98531; Chandrasekhar 1968, EFE) essentially coincide with points $M_3$ and $M_4$. Three type 3 $\lambda$–transitions, leading to various multi-fluid-body sequences, are indicated by small solid dots (see also Table 2). Letter symbols indicate points discussed in the text.

FIGURE 5. *The free energy $E_o$ is plotted versus $j_o^2$ for the sequences of Figure 4. Three type 3 $\lambda$–transitions, leading to various multi–fluid–body sequences, are indicated by small solid dots (see also Table 2). Some letter symbols and labels and the open circles on the quadruple sequence seen in Figure 4 are omitted here for clarity.*

(b) The $\lambda$–transition $R_3 \to S_3 \to T$ in Figures 4 and 5 was determined from the value $j_o^2 = 1.980$ at the lower turning point $R_3$ of the one–ring sequence. An alternative determination, based on the value $j_o^2 = 1.954$ of the starting point $S_3$ of the triple sequence, yields essentially the same answer. The transition leads to a three–fluid–body system (point T in Table 2 and in Figures 4 and 5). On the Maclaurin sequence, the above two $j_o$–values correspond to points $M_3$ and $S_3'$, respectively, that are also listed in Table 2. These points are essentially both identical to the only fourth–harmonic neutral point ($M_3'$ in Table 2) for which Eriguchi & Hachisu (1982) did not find a bifurcating sequence.

(c) The $\lambda$–transition $R_4 \to S_4 \to Q$ in Figures 4 and 5 was determined from the value $j_o^2 = 2.159$ at the starting point $S_4$ of the quadruple sequence. This transition leads to a four–fluid–body system (point Q in Table 2 and in Figures 4 and 5). On the Maclaurin sequence, the above $j_o$–value corresponds to point $M_4$ that is also listed in Table 2. This point is esentially identical to the fourth–harmonic neutral point ($M_4'$ in Table 2) which is also the bifurcation point of the one–ring sequence (see Table 1).

TABLE 3
COMPARISON BETWEEN $\lambda$–TRANSITIONS[a]

| Type | Instability | Spatial Symmetry | Specific Heat | First–Order Transition? | Elementary Catastrophe | Related Sequence | Thermodynamics Analog[b] |
|---|---|---|---|---|---|---|---|
| 1 | Dynamical | Preserved | Continuous | Yes | Cusp | Maclaurin Toroid | Bose–Einstein Condensation |
| 2 | Secular | Preserved | $\infty$ | Yes | (none) | One–ring & Binary | Superfluid Liquid $^4$He |
| 3 | Dynamical | Broken | $\infty$ | Yes | ?[c] | Multi–fluid Bodies | Binary Alloys |
| 4 | Secular | Preserved | $\infty$ | No | Cusp | Two–ring | Ferromagnetism[d] |

(a) Topology breaks and fission occurs in all cases; see also Paper II.
(b) See Huang (1963) for descriptions of the thermodynamical transitions.
(c) If elementary, perhaps then symmetry–restricted butterfly catastrophe; see §9.5 in Poston & Stewart (1978).
(d) See §14.2 in Poston & Stewart (1978).

All the above type 3 $\lambda$–transitions represent dynamical instabilities that lead to breakings of both the symmetry and the topology of the original figure. Related examples from linear stability analyses and from hydrodynamical simulations are discussed in §6 below. The breaking of the symmetry should take place prior to the breaking of the topology. This implies that type 3 $\lambda$–transitions cannot take place before the corresponding critical points are reached because presumably there exist some sequences with branches of unstable equilibria acting as free–energy barriers. Such unstable equilibria do not belong to the unstable branches of the multi–fluid–body sequences which have lower free energies than the mother (Maclaurin and one–ring) sequences (Table 2 and Figure 5). Nonetheless, the coincidence between critical points seen in Figures 4 and 5 suggests some kind of alignment between turning or critical points in the two branches (the multi–fluid–body and the unknown branch) that represent unstable equilibria.

It is not known which sequences act as free–energy barriers of type 3 $\lambda$–transitions and, consequently, whether the barriers disappear smoothly or discontinuously at the $\lambda$–points in Figures 4 and 5. Thus, it is not known whether type 3 dynamical transitions can be described by one of Thom's (1975) elementary catastrophes. If the barriers disappear smoothly, then the so–called "symmetry–restricted butterfly catastrophe" (Poston & Stewart 1978, §9.5; Gilmore 1981, §10.7) is a good candidate for the description of these transitions because its control plane exhibits three distinct regions where the properties of the free–energy function differ. A system evolving along a path that crosses all three regions may then undergo symmetry breaking in the second region and topology breaking in the third region.

A detailed comparison between the various types of hydrodynamical $\lambda$–transitions is provided in Table 3 where we have collected the results of Paper II and of §3 and §4 of this paper. Table 3 summarizes our classification of $\lambda$–transitions and relates them to the equilibrium sequences analyzed in this series of papers as well as to the corresponding thermodynamical phase transitions and, where possible, to elementary catastrophes. The most interesting result contained in this table is that type 2 $\lambda$–transitions are not elementary catastrophes because the free–energy barriers do not disappear smoothly at the corresponding $\lambda$–points, suggesting the absence of degeneracy and the existence of a discontinuity at the critical points of the free–energy functions themselves (see also Paper II).

## 5   Nonlinear resonances

Figure 5 above was meant to illustrate the relative free energies between equilibria at the two ends of type 3 $\lambda$–transitions. However, another interesting effect can also be seen in Figure 5. The lower turning point of the dumbbell–binary sequence and the turning points of the triple and quadruple sequences approach very close to the Maclaurin sequence in the $(j_o^2, E_o)$ plane although they remain at slightly lower free energies relative to the Maclaurin points of the same angular momentum. This kind of structure in the free energies suggests the possibility of *nonlinear resonances*. The three nonlinear resonances of Figure 5 are illustrated also in the $(j_o^2, \omega_o^2)$ plane of Figure 6 by small open circles. The physical properties of each resonance are summarized in Table 4.

<mark>PHASE–TRANSITION THEORY OF INSTABILITIES. IV.</mark> 15

FIGURE 6. *The rotation frequency squared $\omega_o^2$ along the incompressible sequences of Figure 4 is plotted again as a function of the dimensionless angular momentun squared $j_o^2$ but three nonlinear resonances between the Maclaurin sequence and the multi–fluid–body sequences are now indicated by small open circles. The resonances occur at the values $j_o^2$=1.1664, 1.704, and 2.082 of the turning points A, P, and C, respectively, of the multi–fluid–body sequences (see also Table 4). These values were obtained from the corresponding tables published by Eriguchi & Hachisu (1983b) and by Hachisu & Eriguchi (1984a). On the Maclaurin sequence, the resonances appear at points X, Y, and Z that correspond to values of the meridional eccentricity e=0.94455, 0.97343, and 0.98348, respectively. The equilibrium free energies $E_o$ in these resonances may be seen in Figure 5 which shows the turning points of the binary, triple, and quadruple sequences approaching very close to the Maclaurin sequence in the $(j_o^2, E_o)$ plane. Letter symbols indicate points discussed in the text.*

The Maclaurin resonant points X, Y, and Z in Figure 6 and in Table 4 have only slightly higher free energies than the corresponding turning points of the multi–fluid–body sequences. The transitions to the lower energy states are not allowed to occur automatically because each pair of equilibrium states is presumably separated by a free–energy barrier. Such a free–energy barrier must be overcome by finite–amplitude, nonlinear perturbations before a Maclaurin spheroid may break in multiple bodies. So, the resonant points X, Y, and Z in Figure 6 may be thought as critical points of "first–order phase transitions." However, it is not known which equilibrium sequences play the role of the free–energy barriers.

Point X occurs at $j_o^2 = 1.1664$, i.e., prior to the point of second–harmonic dynamical instability on the Maclaurin sequence ($j_o^2 = 1.27957$; Table 1). Technically then, finite–amplitude disturbances can cause fission of the Maclaurin spheroid at X although it is dynamically stable to all infinitesimal perturbations. On the other hand, points Y and Z occur past point M in Figure 6 when the Maclaurin spheroid is already linearly unstable to both second-harmonic and third–harmonic perturbations. A point of fourth–harmonic dynamical instability on the Maclaurin sequence has not been located yet, so we do not know whether it occurs prior to or past point Y or Z in Figure 6.

We see from Figures 5 and 6 that the X $\to$ A resonance leads to the stable branch of the dumbbell–binary sequence, the Y $\to$ P resonance leads to the stable branch of the triple sequence, and the Z $\to$ C resonance leads to the stable branch of the quadruple sequence. Let us now imagine nonlinear evolution driven by finite–amplitude disturbances and taking place at these resonances. Once at the lower turning point of the dumbbell–binary sequence, a system does not necessarily continue to evolve along the binary branch. It may just as well proceed to an even lower energy state on the Jacobi sequence by coalescence and further release of energy (see Figure 5). Similarly, once at the turning points of the triple or the quadruple sequence, a system may eject one or more of its components and make the transition to a state with fewer fluid bodies and lower energy. The ejection mechanism may break the entire multiple system apart to single unbound components but, in some cases, it may also lead to the formation of a bound binary system. In all cases, further evolution beyond the turning points becomes possible because of the strong nonlinear disturbances that have to be present in order to initiate this type of transition at a resonance.

## 6  Discussion and Summary

We have described the phase transitions that appear at higher than second–harmonic critical points on the Maclaurin sequence. These critical points are listed in Table 1. We have found that second–order phase transitions at the third–harmonic bifurcation of the triangle sequence and at the fourth–harmonic bifurcation of the square sequence are forbidden because the daughter sequences exist at higher energies relative to the Maclaurin sequence. Similarly, third–order phase transitions at the third–harmonic bifurcation of the ammonite sequence and at the axisymmetric sixth–harmonic bifurcation of the two–ring sequence are also energetically forbidden. In addition, the ammonite sequence exists only at higher rotation frequencies relative to the Maclaurin sequence and exhibits exactly the same structure as the pear–shaped sequence that bifurcates from the Jacobi sequence (Paper III). The triangle, square, and ammonite sequences terminate prematurely due to equatorial mass shedding while they are still unstable relative to the Maclaurin sequence (see Figure 1 in §2). Since there exist no lower energy branches, no instability of any kind (including $\lambda$–transitions) is associated with the triangle, square, and ammonite sequences. These three sequences of objects in uniform rotation without any allowed phase transition are counter–examples to the common notion that bifurcation points are unconditionally points of secular instability as well. (See also Papers II and III for more related examples and for a discussion of this point.)

The two–ring sequence described in §3 has four turning points that change its structure and dynamical properties. Like the one–ring sequence in Paper II, the two–ring sequence does have a lower energy branch which however is accessible only via a new type of $\lambda$–transition. We call this topology–breaking, symmetry–preserving transition a "$\lambda$–transition of type 4" because, in contrast to the other types, it is not preceded by a first–order phase transition (see Figures 2 and 3 in §3). The type 4 $\lambda$–transition of the two–ring sequence leads on a secular time scale to nonlinear fission to two coaxial axisymmetric toroidal figures that rotate uniformly and with the same frequency.

We have also described two more kinds of nonlinear fission processes, the $\lambda$–transitions of type 3 (§4, Table 2, and Figures 4, 5) and the nonlinear resonances of the Maclaurin sequence (§5, Table 4, and Figures 5, 6). The point of third–harmonic dynamical instability on the Maclaurin sequence lies exactly at the same angular momentum value as the higher turning point of the dumbbell–binary sequence. Thus, this point appears to be a $\lambda$–point. The $\lambda$–transition is illustrated in Figures 4 and 5. It is of type 3 because the symmetry is also broken in addition to the topology. In fact, the symmetry breaks first since the linear dynamical instability is caused by third–harmonic perturbations. The topology breaks second during nonlinear evolution and this $\lambda$–transition finally results in fission of a Maclaurin spheroid to a binary. This type of evolution has been discussed in Paper II and has been observed in the numerical, smoothed–particle hydrodynamical simulations of Miyama, Hayashi, & Narita (1984) during nonequilibrium evolution of a hot isothermal model with $T/|W| = 0.3$. This value is in good agreement with the critical value of $T/|W| = 0.30329$ listed in Table 2.

We have presented in §4 two additional examples of type 3 $\lambda$–transitions that originate on the lower toroidal branch of the one–ring sequence. The $\lambda$–points on the one–ring sequence appear to be related to the starting points of the triple and quadruple sequences, respectively, as well as to two of the three known neutral fourth–harmonic points on the Maclaurin sequence. These relationships are illustrated in Figures 4 and 5 and are quantified in Table 2. It is not known which equilibrium sequences act as free–energy barriers preventing the transitions prior to the $\lambda$–points of type 3.

The physical characteristics of the four types of $\lambda$–transitions are compared in Table 3. The analogies drawn between the hydrodynamical $\lambda$–transitions and the thermodynamical phase transitions and catastrophes are based on the common properties described in §2 and §3 above and in Paper II. However, these analogies are speculative for the most part and need to be confirmed by future work.

The type 3 $\lambda$–transitions from the one–ring sequence lead on a dynamical time scale to fission to three and four detached fluid bodies. These dynamical instabilities are important because their exact counterparts have been discovered by linear stability analyses of (in)–compressible, *differentially rotating*, self–gravitating tori and annuli (this confirms once again the detection of dynamical type 3 $\lambda$–transitions by linear analyses) and have been thoroughly studied in the nonlinear regime with the help of numerical simulations. Such unstable modes were first discovered by Goodman & Narayan (1988) who called them the "Intermediate" or "I–modes." Subsequent linear analyses and numerical simulations of two–dimensional annuli helped to identify the I–modes with nonlinear fission (Christodoulou &

Narayan 1992; Christodoulou 1993). The nonlinear manifestations of the instability were also studied numerically in three–dimensional tori (Tohline & Hachisu 1990; Woodward, Tohline, & Hachisu 1994). Our results leave no doubt that the dynamically unstable I–modes in tori and annuli are responsible for nonlinear fission to multiple fluid bodies and, in addition, they clearly identify these I–modes with type 3 $\lambda$–transitions. Combined with the analytical and the numerical work on I–modes cited above, this identification suggests that the $\lambda$–transitions discussed in this series of papers do occur also in compressible, differentially rotating, self–gravitating fluids and are thus relevant to the evolution of astrophysical fluid systems as well.

We have demonstrated in §5 the power of the nonlinear method that we employ to study instabilities by using the equilibrium free energies shown in Figure 5 to identify nonlinear resonances between the Maclaurin sequence and various multi–fluid–body sequences. Resonances occur at points where the free energies of the binary, triple, and quadruple sequences are only slightly lower than those of the Maclaurin sequence for the same angular momentum value. We have described the physical properties of three such resonances based on Figures 5 and 6 and we have summarized the relevant parameters in Table 4. Resonances are fully nonlinear phenomena and cannot be detected by stability analyses that are accurate to any particular order. Although the free–energy differences are small at resonances (see Figure 5), phase transitions such as those depicted in Figure 6 can only be induced by finite–amplitude disturbances because free–energy barriers presumably stand between the two equilibrium states. Therefore, resonances effectively appear as critical points of "first–order phase transitions" since a lower energy state has opened up to an evolving Maclaurin spheroid but a free–energy barrier prevents the transition from taking place automatically. It is not known which equilibrium sequences form the free–energy barriers.

Of the three nonlinear resonances listed in Table 4 and illustrated in Figure 6, the Maclaurin–to–binary resonance is the most interesting because it appears at $T/|W| = 0.25947$, i.e. prior to the point $T/|W| = 0.27382$ of second–harmonic dynamical instability. This last point is the lowest point of dynamical instability on the Maclaurin sequence predicted by linear analysis. Thus, a Maclaurin spheroid subject to finite–amplitude disturbances at this resonance may break up to a binary although it is dynamically stable to all infinitesimal perturbations.

As we have mentioned in §5 above, because of the extreme nonlinearities involved, dynamical evolution at each resonance does not necessarily have to continue along the binary, triple, or quadruple sequence, respectively, since there exist nearby states of even lower energy. Specifically, a multiple fluid system produced by finite–amplitude disturbances may coalesce forming a Jacobi ellipsoid or it may even eject some/all of its components producing a lower energy system with fewer components or several single unbound bodies (cf. Figures 5 and 6). In some cases, the ejection mechanism may however produce a bound binary, so this process may be of some importance to the formation of binary stars.

Equilibrium sequences of multiple fluid bodies with different compressibilities as well as binary sequences with different mass ratios have been calculated numerically by Eriguchi & Hachisu (1983b, 1984), Hachisu & Eriguchi (1984a, b), and Hachisu (1986) who also discussed briefly some of the phase transitions between sequences. Their results help us understand better the nonlinear fission processes that we have described mainly in this paper and in Papers II and III. Such processes may be responsible for the formation of multiple stellar systems and, in particular, binary stars that seem to be more common than single stars in our Galaxy. [See the reviews of Abt (1978, 1983) and Bodenheimer (1992), the recent work of Abt, Gomez, & Levy (1990), and references therein.] Capture of stars in stellar clusters has been proposed as an alternative mechanism of binary–star formation. It is, however, unlikely that capture may be responsible for the formation of the observed short–period, spectroscopic binaries.

The capture mechanism strives to explain the observed unequal–mass binaries (Abt, Gomez, & Levy 1990). The presumption is that the classical fission mechanism from the Jacobi sequence (Paper III) can only lead to the formation of equal–mass binaries. Although this is true, capture is not the only mechanism that produces unequal–mass binaries. The evolutionary paths described above, and the type 3 $\lambda$–transitions in particular, may also end to unequal–mass binaries. This is particularly easy to achieve via the $\lambda$–transitions that originate from the toroidal states of the one–ring sequence. These transitions may easily form unequal–mass multiple systems especially when the original toroidal structures are unstable to several modes of comparable growth rates (cf. Christodoulou & Narayan 1992). Unequal–mass binaries may then be left behind after the ejection of the additional low–mass components. Therefore, the presumption that unequal–mass binaries can form only by capture is no longer justified in view of such nonlinear fission processes.

But how can *unequal–mass short–period binaries* form? It is difficult to have such systems form by capture and it is not clear whether the above nonlinear fission processes can lead to their formation. The unequal–mass binary sequences lie between the stable parts of the Maclaurin and Jacobi sequences and the equal–mass binary sequence (Hachisu & Eriguchi 1984b; see also Paper III). Perhaps these sequences can be accessed from the stable parts of the Maclaurin and the Jacobi sequences through additional nonlinear resonances but this matter has not yet been explored.

At this point, we close our discussions of results as well as this entire series of papers with an important reminder: Most of our results and conclusions have been obtained by analyzing rotating, self–gravitating, *incompressible fluid* models and cannot be extended by default to the studies of compressible and/or magnetized systems where internal, magnetic, and other forms of energy may play an important role. (Such energy sources have been excluded from the free–energy functions that we have used so far.) The theory of phase transitions is, however, applicable also to *compressible* (see Paper III) and to *stellar* (see Paper I) systems and is expected to yield a clearer physical interpretation of the instabilities, catastrophes, and bifurcations encountered in such systems as well.

## Acknowledgments

We thank C. McKee, and R. Narayan for stimulating discussions and I. Hachisu and N. Lebovitz for useful correspondence. We also thank the main referee of this series of papers for comments that led to improvements in the presentation of results. IS is grateful to the Gauss Foundation for support and to K. Fricke, Director of Universitäts-Sternwarte Göttingen, for hospitality during a stay in which much of this work has been accomplished. IS thanks also the Center for Computational Studies of the University of Kentucky for continuing support. This work was supported in part by NASA grants NAGW–1510, NAGW–2447, NAGW–2376, and NAGW–3839, by NSF grant AST–9008166, and by grants from the San Diego Supercomputer Center and the National Center for Supercomputing Applications.

<div style="text-align:center">REFERENCES</div>

PHASE–TRANSITION THEORY OF INSTABILITIES. IV.                                21Poston, T., & Stewart, I. N. 1978, Catastrophe Theory and its Applications (London: Pitman)
Thom, R. 1975, Structural Stability and Morphogenesis (Reading: Benjamin)
Tohline, J. E. 1985, ApJ, 292, 181
Tohline, J. E., & Christodoulou, D. M. 1988, ApJ, 325, 699
Tohline, J. E., & Hachisu, I. 1990, ApJ, 361, 394
Woodward, J. W., Tohline, J. E., & Hachisu, I. 1994, ApJ, 420, 247

TABLE 2
TYPE 3 $\lambda$–TRANSITIONS

| Point | $j_o^2$ | $\omega_o^2$ | $E_o$ | $T/|W|$ | $e$ |
|---|---|---|---|---|---|
| MACLAURIN TO BINARY SYSTEM: M $\to$ D $\to$ B | | | | | |
| M | 1.540 | 10.48661 | $-7.10825$ | 0.30329 | 0.96705 |
| D | 1.540 | 3.288 | $-7.754$ | 0.18260 | ... |
| B | 1.540 | 0.3594 | $-8.002$ | 0.06700 | ... |
| Third–harmonic Dynamical Instability (EFE) | | | | | |
| M$'$ | 1.53796 | 10.49141 | $-7.10294$ | 0.30308 | 0.96696 |
| ONE–RING TO TRIPLE SYSTEM: R$_3$ $\to$ S$_3$ $\to$ T | | | | | |
| R$_3$ | 1.980 | 7.556 | $-8.031$ | 0.32280 | ... |
| S$_3$ | 1.954 | 3.455 | $-8.430$ | 0.22900 | ... |
| T | 1.980 | 1.044 | $-8.612$ | 0.14230 | ... |
| On the Maclaurin Sequence | | | | | |
| M$_3$ | 1.980 | 9.29812 | $-8.10289$ | 0.34393 | 0.98126 |
| S$_3'$ | 1.954 | 9.37373 | $-8.05183$ | 0.34180 | 0.98065 |
| Fourth–harmonic Neutral Point (EFE) | | | | | |
| M$_3'$ | 1.96741 | 9.33478 | $-8.07828$ | 0.34290 | 0.98097 |
| ONE–RING TO QUADRUPLE SYSTEM: R$_4$ $\to$ S$_4$ $\to$ Q | | | | | |
| R$_4$ | 2.159 | 5.796 | $-8.417$ | 0.31200 | ... |
| S$_4$ | 2.159 | 3.826 | $-8.490$ | 0.26800 | ... |
| Q | 2.159 | 2.324 | $-8.514$ | 0.21960 | ... |
| On the Maclaurin Sequence | | | | | |
| M$_4$ | 2.159 | 8.77090 | $-8.43155$ | 0.35780 | 0.98495 |
| Fourth–harmonic Neutral Point (EFE) | | | | | |
| M$_4'$ | 2.17884 | 8.71210 | $-8.46563$ | 0.35925 | 0.98531 |

TABLE 4
Nonlinear Resonances

| Point | $j_o^2$ | $\omega_o^2$ | $E_o$ | $T/|W|$ | $e$ |
|---|---|---|---|---|---|
| MACLAURIN TO BINARY SYSTEM: X → A | | | | | |
| X | 1.1664 | 11.15073 | −6.00275 | 0.25947 | 0.94455 |
| A | 1.1664 | 1.765 | −6.175 | 0.11800 | ... |
| MACLAURIN TO TRIPLE SYSTEM: Y → P | | | | | |
| Y | 1.704 | 10.07324 | −7.51326 | 0.31964 | 0.97343 |
| P | 1.704 | 2.853 | −7.620 | 0.19800 | ... |
| MACLAURIN TO QUADRUPLE SYSTEM: Z → C | | | | | |
| Z | 2.082 | 8.99873 | −8.29493 | 0.35200 | 0.98348 |
| C | 2.082 | 3.323 | −8.300 | 0.24800 | ... |